\documentclass[10pt, conference]{IEEEtran}
\usepackage{amssymb,amsmath,epsfig}
\usepackage{stmaryrd}
\usepackage{cite}
\usepackage{algorithmic}
\usepackage{algorithm}
\usepackage{psfig}
\usepackage{amssymb}
\usepackage{amsthm}
\usepackage[keeplastbox]{flushend}
\usepackage{multicol}
\usepackage{epstopdf}
\usepackage{lipsum}

\begin{document}
\title{Tracking of the Internal Jugular Vein in Ultrasound Images Using Optical Flow}

\author{\IEEEauthorblockN{Ebrahim~Karami\IEEEauthorrefmark{1},~Mohamed~S.~Shehata\IEEEauthorrefmark{1}, and~Andrew~Smith\IEEEauthorrefmark{2}}
\IEEEauthorblockA{\IEEEauthorrefmark{1}Electrical and Computer Engineering, Memorial University, Canada}
\IEEEauthorblockA{email: \{ekarami,mshehata\}@mun.ca}
\IEEEauthorblockA{\IEEEauthorrefmark{2} Faculty of Medicine, Memorial University, Canada}
\IEEEauthorblockA{email: andrew.smith@med.mun.ca}}

\maketitle
\begin{abstract}
Detection of relative changes in circulating blood volume is important to guide resuscitation and manage variety of medical conditions including sepsis, trauma, dialysis and congestive heart failure. Recent studies have shown that estimates of circulating blood volume can be obtained from ultrasound imagery of the of the internal jugular vein (IJV). However, segmentation and tracking of the IJV is significantly influenced by speckle noise and shadowing which introduce uncertainty in the boundaries of the vessel. In this paper, we investigate the use of optical flow algorithms for segmentation and tracking of the IJV and show that the classical Lucas-Kanade (LK) algorithm provides the best performance among well-known flow tracking algorithms.
\end{abstract}
\section{Introduction}
\begin{figure}
\centering 
\includegraphics[width=1\columnwidth]{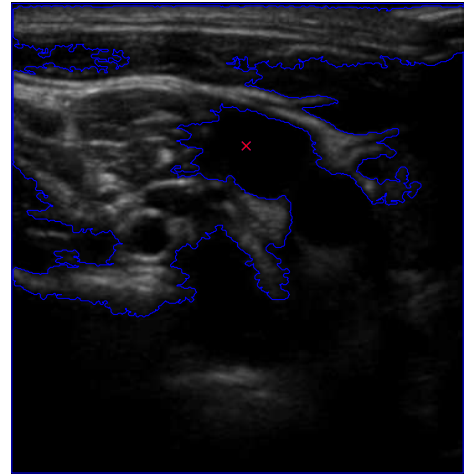}
\vspace{-.5in}
\caption{The result of region growing in the case of broken edge (the red point is the initial seed point).}
\label{RG_broken}
\end{figure}

Detection and monitoring relative changes in circulating blood volume is important for a variety of medical conditions including hemorrhage from trauma and volume overload pertaining to congestive heart failure \cite{sawka2000, yashiro2003,Bremer2004}. Recent studies suggest that the cross-sectional area (CSA) of the internal jugular vein (IJV) can be used to detect and monitor relative changes in blood volume \cite{bailey2012,raksamani2014}. 
\noindent Manual segmentation of IJV from hundreds of ultrasound frames is a time-consuming task making it inappropriate for real-time blood-volume monitoring application. In \cite{qian2014}, the combination of speckle tracking \cite{voigt2014definitions} and active contour (STAC) was proposed for the segmentation and tracking of the IJV in which the coarse segmentation obtained from speckle tracking is smoothed with an active contour. Unfortunately, speckle tracking fails when the IJV undergoes fast variations requiring an ultrasound machine with a high frame rate. This problem was later addressed by cascading region growing with active contour (RGAC) \cite{karami2016}with the main problem being leakage when a section of the vessel wall was obscured as shown in Fig. \ref{RG_broken}.
\noindent Both STAC and RGAC algorithms are based on active contours which deals with IJV contours as a set of points, while the boundaries of the IJV are fuzzy. In this paper, we investigate the utility of several optical flow algorithms including Lucas-Kanade (LK) \cite{lucas1981}, Horn-Schunck (HS) \cite{horn1981}, and Farneback (FB) \cite{farneback2003} in tracking and segmenting the IJV in ultrasound video. 

\indent The paper is organized so that section II introduces three popular optical flow algorithms, investigated in this paper; section III demonstrates the results of segmentation and tracking as applied to the IJV with results compared to \cite{karami2016} and \cite{qian2014}; Section IV summaries the findings.
\section{Optical Flow}
Optical flow is defined as a vector field that relates each frame to the next frame. Assuming that the intensity levels between two frames is preserved, we can write,
\vspace{2mm}
\begin{equation}
I(x,y,t)=I(x+dx,y+dy,t),\label{eq_fl1}
\end{equation}
\vspace{2mm}
\noindent where $I(x,y)$ is the intensity of $t$th frame at pixel $(x,y)$. If the displacement $(dx,dy)$ is very small, then using the taylor series approximation, (\ref{eq_fl1}) can be rewritten as
\vspace{2mm}
\begin{equation}
I_x u +I_y v +I_t=0,\label{eq_fl2}
\end{equation}
\vspace{2mm}
\noindent where $u=dx$, $v=dy$, $I_x=\frac{\partial I(x,y,t)}{\partial x}$, $I_y=\frac{\partial I(x,y,t)}{\partial y}$, and $I_t=I(x,y,t+1)-I(x,y,t)$. The approximation in (\ref{eq_fl2}) is valid only if displacement from one frame to the next frame is very small. For relatively large displacements, optical flow algorithms are generally implemented as coarse-to-fine using a pyramid structure. In this paper, we assume three layer pyramid structures for each of the investigated optical flow algorithms.\\
\subsection{Lucas-Kanade (LK) Algorithm}
The LK algorithm is the most popular optical flow algorithm which has found many applications in computer vision and image processing \cite{lucas1981, karamanidis2016, plyer2016, oron2014}. The LK algorithm solves the optical flow constraint (\ref{eq_fl2}) by dividing the image into smaller blocks and assumes that the displacement of pixels in each block is constant \cite{lucas1981}. Therefore, the constraint (\ref{eq_fl2}) is solved by minimizing the following energy function:
\vspace{2mm}
\begin{equation}
E(u,v)=\sum\limits_{x,y \in \Omega}W(I_x u+I_y v +I_t)^2, \label{eq_fl3}
\end{equation}
\vspace{2mm}
\noindent where $\Omega$ is the block around the pixel and $W$ is the weights given to the elements such that more emphasis is placed on pixels near the center of each block. The solution for minimizing (\ref{eq_fl3}) is obtained as
\vspace{2mm}
\begin{equation}
\left[\begin{array}{ll} \sum W I_x^2 & \sum W I_x I_y \\ \sum W I_x I_y & \sum W I_y^2\end{array}\right]\left[\begin{array}{c} u \\ v \end{array} \right] = -\left[\begin{array}{c} \sum W I_x I_t \\ \sum W I_y I_t \end{array} \right].
\end{equation}
\subsection{Horn-Schunck (HS) Algorithm}
The HS algorithm \cite{horn1981} computes an estimate of the displacement vector $[u v]^T$ by assuming that the flow vector field is smooth over the entire image \cite{horn1981, meinhardt2013}. Therefore, the energy function is modified as
\vspace{2mm}
\begin{equation}
\begin{split}
&E(u,v)=\int\int(I_x u+I_y v +I_t)^2 dx dy \label{hs_1}\\&+\alpha \int\int \left\{ (\frac{\partial u}{\partial x})^2+(\frac{\partial u}{\partial y})^2+(\frac{\partial v}{\partial x})^2+(\frac{\partial v}{\partial y})^2\right\} dx dy,
\end{split}
\end{equation}
\vspace{2mm}
\noindent where $\alpha$ is an scale factor for the last term forcing the algorithm to provide smooth displacement over the image. This minimization is obtained by the following iterative equations:
\begin{equation}
u_{x,y}^{m+1}=\bar{u}_{x,y}^{m}-\frac{I_x^2\bar{u}_{x,y}^{m}+I_x I_y\bar{v}_{x,y}^{m}+I_x I_t}{\alpha^2+I_x^2+I_y^2},
\end{equation}
\begin{equation}
v_{x,y}^{m+1}=\bar{v}_{x,y}^{m}-\frac{I_x I_y\bar{u}_{x,y}^{m}+I_y^2\bar{v}_{x,y}^{m}+I_y I_t}{\alpha^2+I_x^2+I_y^2},
\end{equation}
\noindent where $u_{x,y}^{m}$ and $v_{x,y}^{m+1}$ are horizontal and vertical displacement of pixel $(x,y)$ estimated at $m$th iteration, respectively, and $\bar{u}_{x,y}^{m}$ and $\bar{v}_{x,y}^{m+1}$  are their corresponding neighborhood average.
\par
\subsection{Farneback (FB) Algorithm}
Both LK and HS optical flow algorithm assume that the first order Taylor approximation is sufficient for motion tracking. The idea of the FB algorithm is based on approximating the neighborhood of the pixel $x$ of the $i$th frame with a quadratic polynomial as follows \cite{farneback2003}
\vspace{2mm}
\begin{equation}
f_i(x)=\mathbf{x^T A_i x+b_i^Tx}+c_i,\label{eq:FB1}
\end{equation}
\vspace{2mm}
\noindent where $x$ is a $2\times 1$ vector corresponding to the pixel coordination, $\mathbf{A_i}$ is a $2\times2$ symmetric matrix, $\mathbf{b_i}$ is a $2\times1$ vector, $c_i$ is an scalar, and superscript $T$ denotes a transpose operation. A displacement $d$ is estimated as 
\vspace{2mm}
\begin{equation}
    f_2(x)=f_1(x-d).\label{eq:FB2}
\end{equation}
By substitution of (\ref{eq:FB2}) in (\ref{eq:FB1}), $d$ is estimated as
\begin{equation}
    d=-\frac{1}{2}\mathbf{A_1^{-1}(b_2-b_1)}.\label{eq:FB3}
\end{equation}
\vspace{3mm}
\section{Results}
Experimental ultrasound video clips of the IJV were collected from 14 healthy subjects at a variety of inclinations with the head of the bed elevated at 0, 30, 45, 60, and 90 degrees designed to simulate relative changes in blood volume. The IJV was imaged in the transverse plane using a portable ultrasound machine (M-Turbo, Sonosite-FujiFilm) and a linear-array probe (6-15 Mhz). Each video had a frame rate of 30 fps, scan depth of 6cm, and a duration of 15 seconds (450 frames/clip). The three optical flow algorithms were compared with expert manual segmentation, region-growing-based-active-contour (RGAC) \cite{karami2016} and speckle-tracking-based-active-contour (STAC) algorithms \cite{qian2014}. The number of contour points is assumed to be $N=32$ in all algorithms. The number of pyramid layers and window length in the optical flow algorithms was set at 3 and 20 pixels, respectively. The maximum length for RGAC algorithm was set to 70 pixels. For the HS algorithm, the smoothing factor, $\alpha$, was set to one and the algorithm run for 250 iterations. The DICE coefficient, also known as Sorensen index, was used to determine the level of agreement between each algorithm and manual segmentation results \cite{dice1945measures, sorensen1948method}. The DICE factor $S$ is defined as:
\begin{equation}
S=\frac{2|A\cap M|}{|A|+|M|},
\end{equation} 
\noindent where $|A|$ and $|M|$ are the CSA of the IJV estimated from the algorithms and the manual segmentation, respectively, and $|A\cap M|$ the intersection of the area between them.\par
We define a failure as having a DICE co-efficient below 0.7 during IJV segmentation and tracking. By this definition, the number of cases (among 70 videos), where RGAC, STAC, LK, HS, and FB algorithms successfully track and segment the IJV are 4, 9, 31, 21, and 26 videos, respectively. This demonstrates the ability of optical flow algorithms to successfully segment and track the IJV and highlights their success rate when compared to RGAC and STAC algorithms.
\par
\begin{figure}[t!]
\centering
\includegraphics[width=1\linewidth]{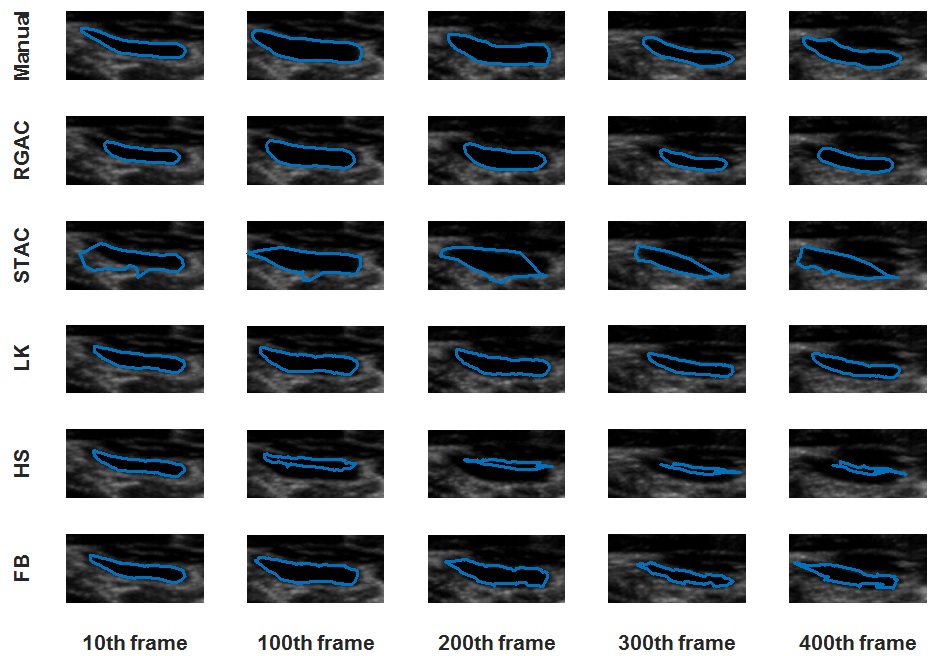}

\caption{Segmentation and tracking Performance of all five algorithms on a good quality video.}
\label{Table8_60}
\end{figure}
\begin{figure}[t!]
\centering
\includegraphics[width=1\linewidth]{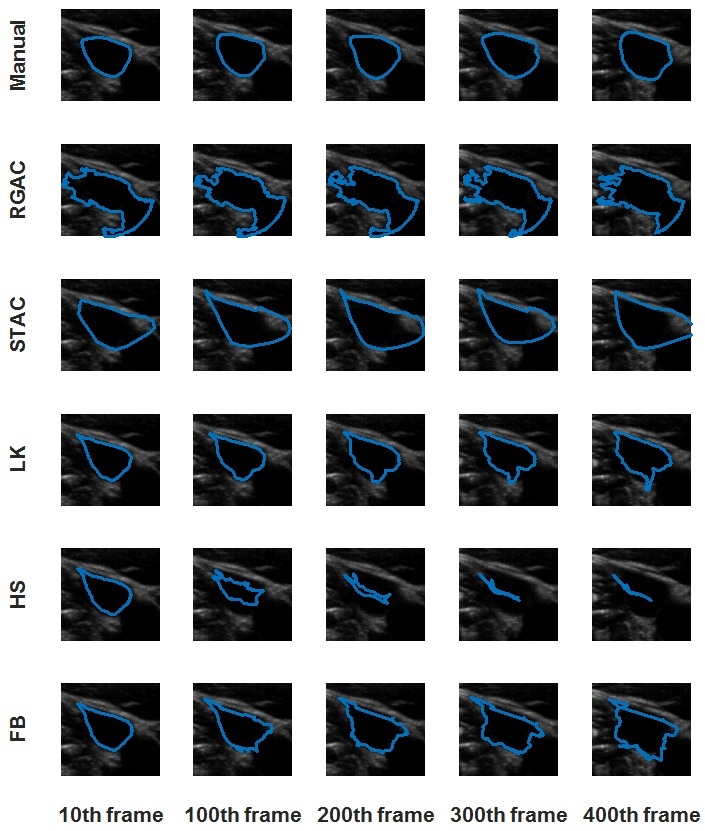}
\caption{Segmentation and tracking Performance of all five algorithms on a poor quality video (with partially missing boundaries).}
\label{Table4_0}
\end{figure}

Figs. \ref{Table8_60} and \ref{Table4_0} detail the tracking performance of the algorithms compared with manual segmentation on a good and poor quality video, respectively. It is evident from Fig. \ref{Table8_60}, both STAC and HS algorithms gradually lose track while the other three algorithms maintain tracking over time. In \ref{Table4_0}, one can see that the RGAC algorithm loses track as a result of leaking through the obscured part of the vessel wall. Both LK and FB algorithms present acceptable performance in this scenario.
\par
\begin{figure}
\centering
\includegraphics[width=1\linewidth]{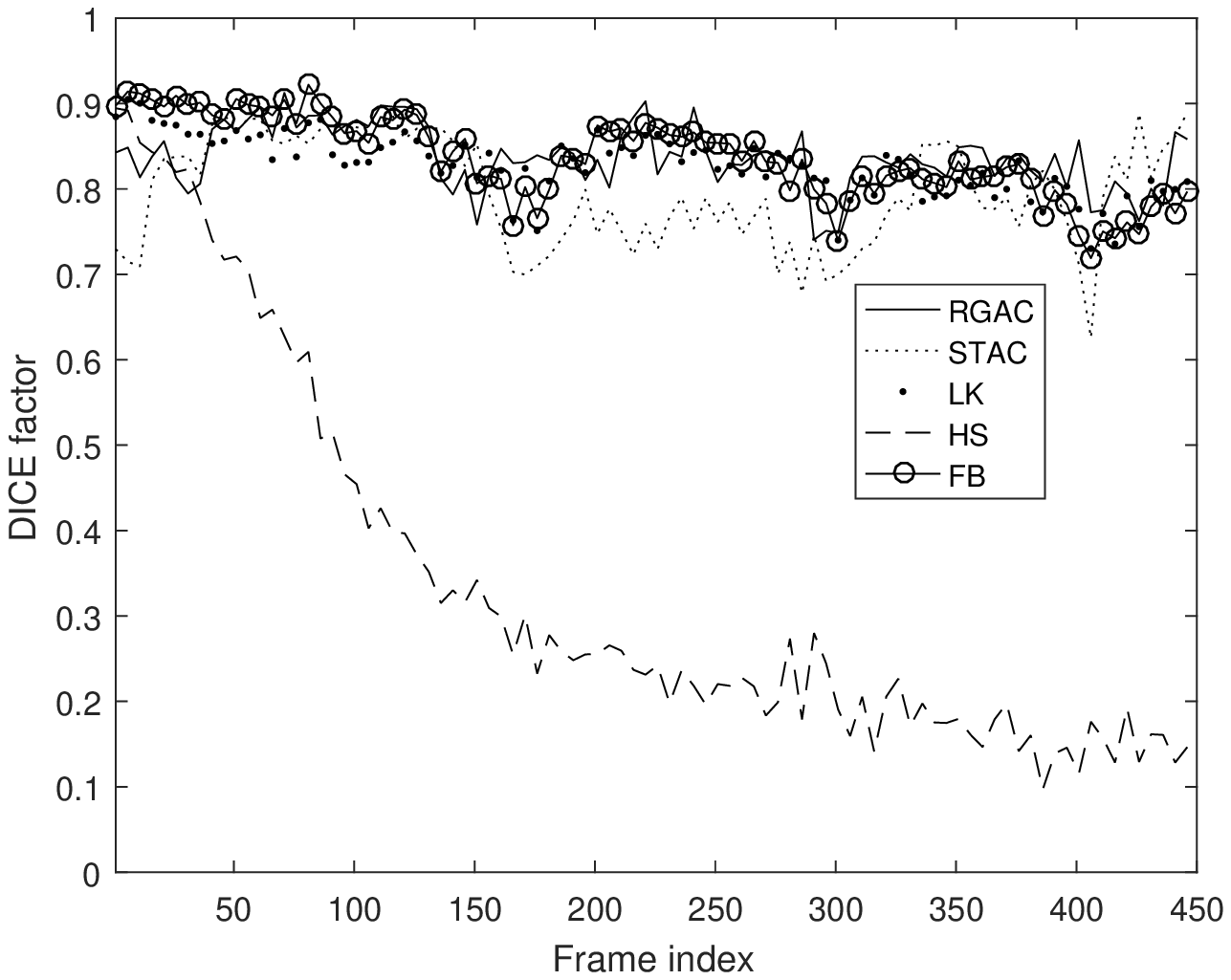}
\caption{The level of agreement between manual segmentation and algorithms versus frame index on a good quality video.}
\label{Dice8_60}
\end{figure}
\begin{figure}
\centering
\includegraphics[width=1\linewidth]{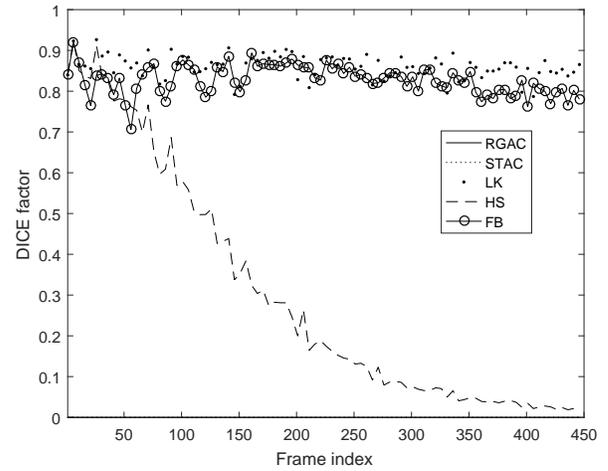}
\caption{The level of agreement between manual segmentation and algorithms versus frame index on a poor quality video.}
\label{Dice4_0}
\end{figure}

Figs. \ref{Dice8_60} and \ref{Dice4_0} highlight the algorithms performance for the two videos considered in Figs. \ref{Table8_60} and \ref{Table4_0}. In both scenarios, it can be seen that the LK method provides the highest DICE factor while the FB algorithm provides the second best performance.

\begin{figure}[t!]
\centering
\includegraphics[width=1\linewidth]{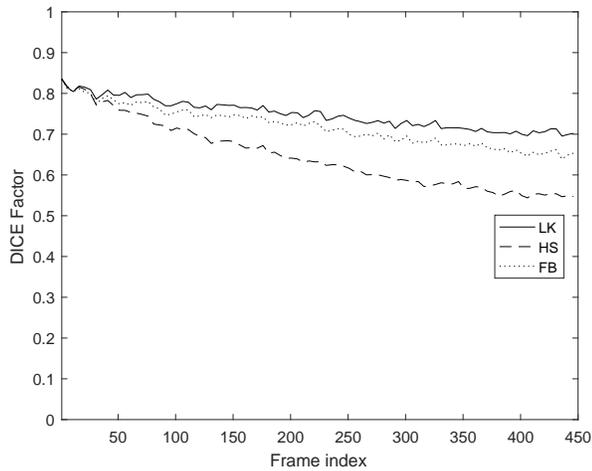}
\caption{The mean level of agreement between manual segmentation and algorithms across all 70 video clips.}
\label{Average_all}
\end{figure}

\begin{figure}
\centering
\includegraphics[width=1\linewidth]{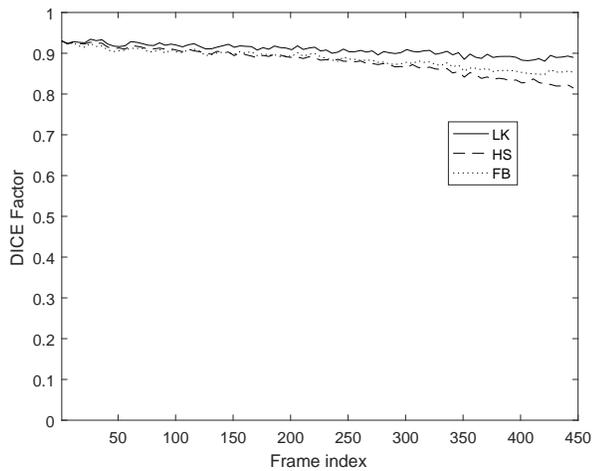}
\caption{The mean level of agreement between manual segmentation and optical flow algorithms for videos (21) in which good tracking has been demonstrated.}
\label{Average_sel}
\end{figure}

Fig. \ref{Average_all} shows the mean level of agreement of the three investigated optical flow algorithms across all 70 IJV videos. Note that each algorithm has failed to properly segment and track the IJV in some of these scenarios resulting in a lower mean level of agreement - particularly evident with HS algorithm. Still, it is evident that the LK algorithm continues to outperform the other optical flow techniques. Fig. \ref{Average_sel} illustrates much better results when the DICE coefficient is averaged over the 21 videos in which the HS algorithm demonstrates successful tracking. In this figure, the LK algorithm continues to provide the best performance. 
\section{Conclusion and Future Work}
In this paper, the utility of optical flow algorithms in tracking and segmenting the internal jugular vein (IJV) in ultrasound videos is investigated. Three optical flow algorithms, namely Lucas-Kanade (LK), Horn-Schunck (HS), and Farneback (FB), were applied to track $N=32$ contour points on the IJV. These algorithms were compared with expert manual segmentation and two existing active-contour-based algorithms previously proposed. The DICE coefficient was used to demonstrate that the optical flow-based algorithms provide good level of agreement with manual segmentation and lower failure rates when compared to previously proposed algorithms.\\
Although the optical flow algorithms outperform the previously published techniques, regular failures to track and segment are present when the boundaries of the IJV are obscured.

\end{document}